\documentclass[twocolumn, superscriptaddress, amsmath, amssymb, aps, doublefloatfix, pra]{revtex4-2}

\usepackage[pdftex]{graphicx}
\usepackage{dcolumn}
\usepackage{bm}
\usepackage{dsfont}
\usepackage{physics}
\usepackage{mathrsfs}
\usepackage[colorlinks, breaklinks]{hyperref}
\usepackage{amsmath}
\usepackage[english]{babel}
\usepackage{booktabs}
\usepackage{amssymb}
\usepackage{type1cm}
\usepackage[caption=false]{subfig}
\usepackage{braket}
\usepackage{xcolor}
\usepackage{comment}

\begin{document}

\title{Quantum Entanglement Distribution via Uplink Satellite Channels}

\author{S. Srikara}
\affiliation{Center for Quantum Software and Information, University of Technology Sydney, NSW 2007, Australia.}

\author{Hudson Leone}
\affiliation{Center for Quantum Software and Information, University of Technology Sydney, NSW 2007, Australia.}

\author{Alexander S. Solnstev}
\affiliation{School of Mathematical and Physical Sciences, University of Technology Sydney, Ultimo, NSW, 2007, Australia.}


\author{Simon J. Devitt}
\affiliation{Center for Quantum Software and Information, University of Technology Sydney, NSW 2007, Australia.}
\affiliation{InstituteQ, Aalto University, 02150 Espoo, Finland.}

\begin{abstract}
Significant work has been done to develop quantum satellites, which generate entangled pairs in space and distribute them to ground stations separated some distance away.
The reverse ``uplink" case, where pairs are generated on the ground and swapped on the satellite using an optical Bell measurement, has not been seriously considered due to a prevailing assumption that it is practically infeasible.
In this letter, we illustrate the feasibility of performing Discrete Variable photonic Bell-measurements in space
by conducting a detailed numerical analysis to estimate the channel efficiency and attainable pair fidelity for various satellite-station configurations.
Our model accounts for a wide range of physical effects such as atmospheric effects, stray photons, and mode-mismatch. 
Our findings show promise toward the feasibility of photonic Bell-measurements in space, which motivates future research towards large-scale Satellite-based uplink entanglement distribution.
\end{abstract}

\maketitle

\section{Introduction}

Distributed entanglement is a crucial resource for quantum communications and distributed quantum computing \cite{quantuminternet, nielsen2001quantum, entanglement1,wehner2018quantum,kimble2008quantum,cacciapuoti2019quantum, cacciapuoti2024multipartite,caleffi2024distributed}. Using satellites to generate and distribute entanglement to distant ground stations has previously been considered in the contexts of global communication networks \cite{downlink1} and resource estimation for fault-tolerant distributed quantum computing \cite{hudsondownlink}.
In particular, the latter work demonstrated that the limited power available to quantum satellites sets a hard limit on the scalability of satellite-based distributed quantum computing in the \textit{downlink} case. 
One potential way to improve scalability is to consider the reverse \textit{uplink} case where the Bell pairs are generated on the ground and then sent to the satellite where they are swapped via Bell measurement. 
On the ground, we can allocate significantly more power to pair production (From MWs to GWs \cite{power_stns}) than we can on a satellite (Around $10$kW \cite{satpower}). This has the potential to not only enhance Bell-pair generation rates significantly, but to also make the satellite lighter and simpler in design \cite{neumann2018q3sat}.

The idea of transferring quantum states from ground to satellite (i.e. via uplink channels) has been previously explored both theoretically and experimentally in both discrete and continuous variables largely for QKD and quantum teleportation \cite{villasenor2021enhanced,zuo2021overcoming,kerstel2018nanobob,neumann2018q3sat,pirandola2021limits,pirandola2021satellite,bonato2009feasibility,behera2024estimating}. A discussion on dual channel uplink transmission for entanglement swapping has been made for continuous variable states \cite{hosseinidehaj2018satellite} but a corresponding discussion with discrete variable photonic states for designing satellite-based repeaters has not yet been made prior to this article because several disadvantages, despite the power advantage, have dissuaded serious consideration of this method.

The first is that uplink channel attenuation is larger than downlink attenuation since atmospheric scattering occurs at the start of the transmission, unlike downlink where it happens in the end \cite{bonato2009feasibility,pirandola2021satellite,maharjan2022atmospheric}.
Additionally, the aperture size of a satellite's receiving telescope is considerably smaller than what is possible on the ground. Due to this, the uplink protocol requires high-precision clock synchronization and satellite positioning \cite{satclocksync} for precisely pointing and aiming the photons at a moving satellite and facilitating the near-simultaneous arrival of the two photons required for a successful Bell measurement.
The above disadvantages pose serious questions on the feasibility of performing successful Bell measurements on the satellite using ground-to-satellite links. Therefore, as a first step towards answering the bigger question of whether the advantages of large power stations on the ground can overpower the disadvantages discussed above, we first need to demonstrate the feasibility of performing Bell measurement on the satellite. 
In this paper, we demonstrate that the uplink Bell measurements are indeed feasible.
This is a first step towards considering realizable practical entanglement distribution with uplink satellite links.

We first introduce our uplink setup and the various physical phenomena affecting the Bell measurement, such as beam widening, beam wandering, atmospheric turbulence and attenuation, stray photons, and mode-mismatch. We ignore polarization error as it is less than 0.001 radians (0.06\%) \cite{bonato2006influence, fedrizzi2009high}, which is negligible. We also ignore the Doppler effect \cite{rouzegar2019estimation,gruneisen2021adaptive} as it can be compensated by precisely measuring the potential wavelength shift based on the trajectory and calibrating the photon wavelength accordingly \cite{liu1999doppler}. Then, we combine all these phenomena to derive expressions for Fidelity and Efficiency of the protocol. Next, we compare these metrics against various important parameters, for which we use a realistic range of values. Lastly, we discuss our results and potential future directions towards the realization of uplink satellite networks. A detailed derivation of the Fidelity and Efficiency expressions can be found in the supplementary material, which, along with the Mathematica code containing the simulations, can be found here: \cite{github}.

\section{The Protocol}\label{sec:setup}

\begin{figure}[b]
\includegraphics[width=\linewidth]{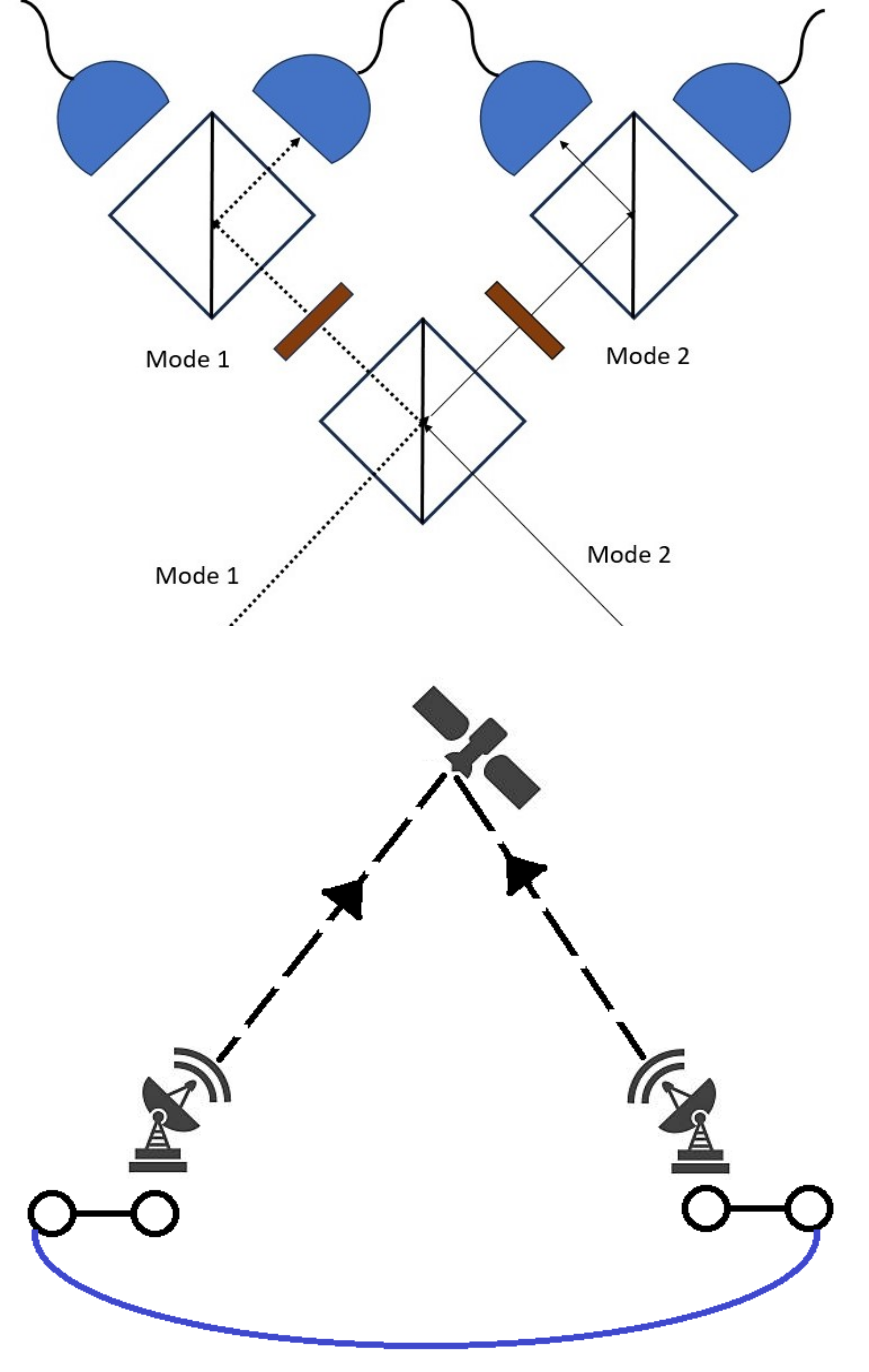}
\caption{\label{fig:schematic}(below) A schematic of the the proposed uplink setup. The setup consists of two ground stations sending one-half of their Bell pairs to a satellite via an uplink channel. The Bell measurement happens on the satellite, entangling the other two ground photons (represented by the dark blue curvy line). (above) The Bell-measurement apparatus inside the satellite. The apparatus consists of two optical modes, directed by a Polarizing Beam Splitter (PBS), followed by two 45$^\circ$ polarizers (one on each mode), and further followed by two PBSes, one on each mode, and foue bucket detectors, one on each output of the final PBSes.}
\end{figure}

{\em Setup.} Our setup consists of two ground stations separated by a distance $D_G$ with a satellite orbiting overhead in the Low-Earth-Orbit (LEO) region (Figure  \ref{fig:schematic}).
For our setup, we fix the satellite equidistant from the two ground stations.
At each ground station, we consider a Bell-pair generator that generates perfect Bell-pairs i.e. with a fidelity of 1. 
One-half of this pair is encoded in the polarisation basis of a photon, and the other half is encoded in a physical qubit of an arbitrary quantum computing architecture installed at the ground station.
The photons at each station are fired up towards the satellite, where they will be entanglement-swapped via optical Bell measurement (Figure \ref{fig:schematic} inset).
This is implemented with the use of a polarizing beamsplitter (PBS) followed by an optical Hadamard (i.e. a 45$^{\circ}$ waveplate) and two polarization-resolving photon detectors.
Each of these detectors themselves are made of a polarizing beamsplitter followed by two photodetectors.
A successful outcome is determined when the satellite detects two clicks -- one at each mode.
This entangles the two computational qubits at the ground stations.

\section{Mode-matching model}\label{sec:modematch}
{\em Mode-Mismatch.} Every photon has probability a wavepacket associated with it. 
During optical Bell measurements, the two incoming photons must arrive at the satellite simultaneously i.e. the wavepackets of the two incoming photons should spatiotemporally align. 
We call this mode-matching \cite{rohde2005frequency,rohde2011time}. 
Any spatiotemporal mismatch between the two wavepackets leads to increased distinguishability between the two photons, thus leading to reduced fidelity of the swapped Bell pairs on the ground.
To reduce the effect of mode-mismatch, we employ time-gating i.e. a short, limited time window when the photon detectors are switched on. 
This allows only a fraction of the wavepackets to be detected, thus decreasing their distinguishability and reducing the probability of successful measurement. 
Therefore, the choice of the length of the gating window depends upon the acceptable tradeoff between fidelity and success-probability of the measurement.

A photon in free space located at position $x_0$ is described by the Gaussian wavepacket $\psi$ centred at $x_0$, given by \cite{essential}:
\begin{align}
    \psi(x) = \left( \frac{1}{2 \pi \sigma ^2} \right) ^\frac{1}{4} e^{-\frac{1}{4} \left( \frac{x-x_0}{\sigma} \right)^2} e^{ik(x-x_0)}, \; \sigma = c \sigma_t
\end{align}
where $\sigma_t$ is the temporal width of the wavepacket \cite{wavepacket_width1,wavepacket_width2} (i.e.\ the standard deviation of the Gaussian), $k=2 \pi / \lambda$ is the wave number (where $\lambda$ is the wavelength of the ground photon), and $c$ is the speed of light.
Consider the two incoming photons from the two ground stations whose associated wavepackets are $\psi_1$ and $\psi_2$ respectively. 
Let's say these photons are incident on the satellite at times $t_1$ and $t_2$. This leads to a temporal mismatch $\Delta t = t_2 - t_1$ between the two photons, which can also be interpreted as the path difference $\Delta x$, such that $\Delta x = c \Delta t$ and therefore $\psi_2(x) = \psi_1(x + \Delta x)$.
For the moment, let's assume that we have an ideal photon channel with mode-mismatch being the only phenomenon affecting our protocol.
Let $t_\mathrm{min}$ and $t_\mathrm{max}$ be the opening and closing time of the gating window i.e. the switch-on time and the switch-off time of the detector respectively.
Then, the probability $P_{gw}$ that the photon entering mode $i$ passes through the gating window is given by [See Supplementary Material Section 4]:
\begin{align}\label{eqn:uplink_modematch_success_prob}
    P_{gw_i} = \int_{c t_\mathrm{min}}^{c t_\mathrm{max}} |\psi_i(x)|^2 dx 
\end{align}
The fidelity $F_{ic}$ of the resulting swapped Bell-pair is given by:
\begin{align}\label{uplink_modematch_fidelity}
    F_{ic} &= \frac{1}{2} + \frac{\left| \int_{c t_\mathrm{min}}^{c t_\mathrm{max}} \psi_1(x)\psi_2^*(x)dx \right|^2}{2 P_{gw_1} P_{gw_2}}
\end{align}

\section{Errors Due to Beam Widening and Wandering}\label{subsec:beam_wandering}

{\em Beam widening and beam wandering.} Beam widening and beam wandering are significant phenomena affecting efficiency in satellite communications. \cite{pirandola2021satellite,bonato2009feasibility,uplink_beam_wandering,esposito1967power,fried1973statistics,titterton1973power,vasylyev2012toward}.
Beam widening is an intrinsic phenomenon where the width of the beam gradually increases with travel, making the overall beam conical \cite{beamwidening1,beamwidening2,beamwidening3,beamwidening4}, while beam wandering is a phenomenon where the beam centroid randomly and continuously shifts as it passes through the atmosphere due to changing refractive index of the atmosphere caused by turbulence \cite{beamWandering,pirandola2021satellite,uplink_beam_wandering}.
If $R_A$ is the radius of aperture of the receiving telescope, and $w$ is the width of the beam due to the combined effects of beam widening and wandering (not to be confused with the width $\sigma$ of the probability wavepacket of the photons), then the channel efficiency $\eta_w$ due to beam widening and wandering is given by \cite{pirandola2021satellite,bonato2009feasibility,bourgoin2013comprehensive}:
\begin{align}\label{eqn:eta_w}
    \eta_w = 1-e^{-2R_{A} ^2/(w^2 + 10^{-12} z^2 + \sigma^2 _{\mathrm{tr}})}
\end{align}
where $10^{-12} z^2$ is the pointing error of the transmitter \cite{pirandola2021satellite} and $\sigma_{\mathrm{tr}}$ is the tracking error in Satellite's position, and $w$ is the overall long-term beam width due to the combined effects of beam widening and wandering \cite{pirandola2021satellite} [See Supplementary Material Section 5.2].

\section{Atmospheric Attenuation}

{\em Atmospheric Attenuation.} The channel efficiency $\eta_{a}$ resulting solely due to atmospheric attenuation is given by \cite{pirandola2021satellite}:
\begin{align}
    \eta_a (h,\theta) = \exp \left[-\alpha_0 \int_{0}^{z(h,\theta)} exp\left[ -\frac{h(y,\theta)}{\Tilde{h}} \right] dy\right]
\end{align}
where $h$ is the satellite altitude, $\alpha_0 = 5 \times 10^{-6} m^{-1}$ and $\tilde{h} = 6600 m$ are constants \cite{atmosphere1,atmosphere2}.

\section{Stray Photons and Noise}\label{sec:stray_ph}

{\em Stray Photons.} There are various sources of stray photon we need to account for in our analysis \cite{pirandola2021limits,leinert19981997,bonato2009feasibility,strayPhotonsDownlink,hansen1984spectral,liorni2019satellite}.
Earth's surface emits numerous stray photons into space from its reflected sunlight in the daytime and from reflected moonlight and Earth's own blackbody radiation in the nighttime \cite{bonato2009feasibility,pirandola2021satellite,strayPhotonsDownlink}. These stray photons, when incident on the satellite along with the legitimate photons, can lead to corrupted measurements.
Let $r_{\mathrm{day}}$ and $r_{\mathrm{night}}$ be the rates of stray photons incident on the photodetectors during daytime and night time respectively. Within a time-gating window $t$ of the detector (where $t = t_{\mathrm{max}} - t_{\mathrm{min}}$)(not to be confused with $\Delta t$ which is the path difference between the two photonic wavepackets divided by $c$), the probability $P_{SP}$ that $n$ stray photons are incident on a single detector follows the Poissonian distribution:
\begin{align}
    P_{SP} (n) = \frac{(r_{\mathrm{day/night}} t)^n e^{-r_{\mathrm{day/night}} t}}{n!}
\end{align}
The detailed derivation of $r_{\mathrm{day}}$ and $r_{\mathrm{night}}$ is provided in Section 6 of the supplementary material.

\section{Dual Uplink Channel Efficiency}\label{sec:overall}

{\em Overall Success Probability.} In our measurement setup (Figure \ref{fig:schematic}), we have four detectors 1, 2, 3, and 4. 
Let the tuple $d = (d_1, d_2, d_3, d_4)$ denote the detection outcomes, with $d_i = \{0, 1\}$ for detector $i$, where $0$ represents no click whereas $1$ represents a click.
Here $d_1$ and $d_2$ represent left mode, and $d_3$ and $d_4$ represent right hand modes.
The probability that photons from the ground stations cause a detection event according to tuple $d$ is $P_G(d)$.
The probability that stray photons will cause detector clicks according to tuple $d$ is $P_D(d)$.
$P_G(d)$ and $P_D(d)$ incorporate all the losses described above (see supplementary material).
The measurement signature $M$ is given by the bit-wise AND of bit-strings associated with $G$ and $D$.
The allowed accepted signatures are $M = (1,0,1,0)$, $(1,0,0,1)$, $(0,1,1,0)$, and $(0,1,0,1)$.
These are legitimate signatures because there is one photon detected in each of the spatially resolved modes.
Consider one success signature $M = (1, 0, 1, 0)$. The probability $P_M$ of this happening is [See Supplementary Material Section 7]:
\begin{align}
    P_M(1, 0, 1, 0) &= P_G(1, 0, 1, 0)[P_D(0, 0, 0, 0) + P_D(0, 0, 1, 0) \nonumber\\ &+ P_D(1, 0, 0, 0) + P_D(1, 0, 1, 0)] \nonumber\\
&+ P_G(0, 0, 1, 0)[P_D(1, 0, 0, 0) + P_D(1, 0, 1, 0)] \nonumber\\
&+ P_G(1, 0, 0, 0)[P_D(0, 0, 1, 0) + P_D(1, 0, 1, 0)] \nonumber\\
&+ P_G(0, 0, 0, 0)P_D(1, 0, 1, 0)
\end{align}
Assuming that the four detectors are equally configured, it can be verified that all four legitimate signatures are symmetric.
Therefore, the total probability $\eta_{tot}$ of successful signature for any given photon pair coming from the ground stations (or equivalently the uplink dual-channel efficiency) is given by:
\begin{align}\label{eqn:eta_tot}
    \eta_{tot} = 4 P_M(1, 0, 1, 0)
\end{align}

\section{Practical Fidelity}\label{sec:final_fid}

{\em Overall Fidelity.} Let the probability of a legitimate coincidence given the success signature M be $P_{S}$. 
In case of an illegitimate coincidence, the resulting state is considered to be maximally mixed ($I/4$). 
Therefore, the actual fidelity $F$ of the resulting swapped Bell-pair is given by [See Supplementary Material Section 8]:
\begin{align}\label{eqn:final_fidelity}
    F = P_S F_{ic} + (1 - P_S)/4
\end{align}

\section{Results and discussion}\label{sec:results}

\begin{figure}
    \centering
    \includegraphics[width=0.9\linewidth]{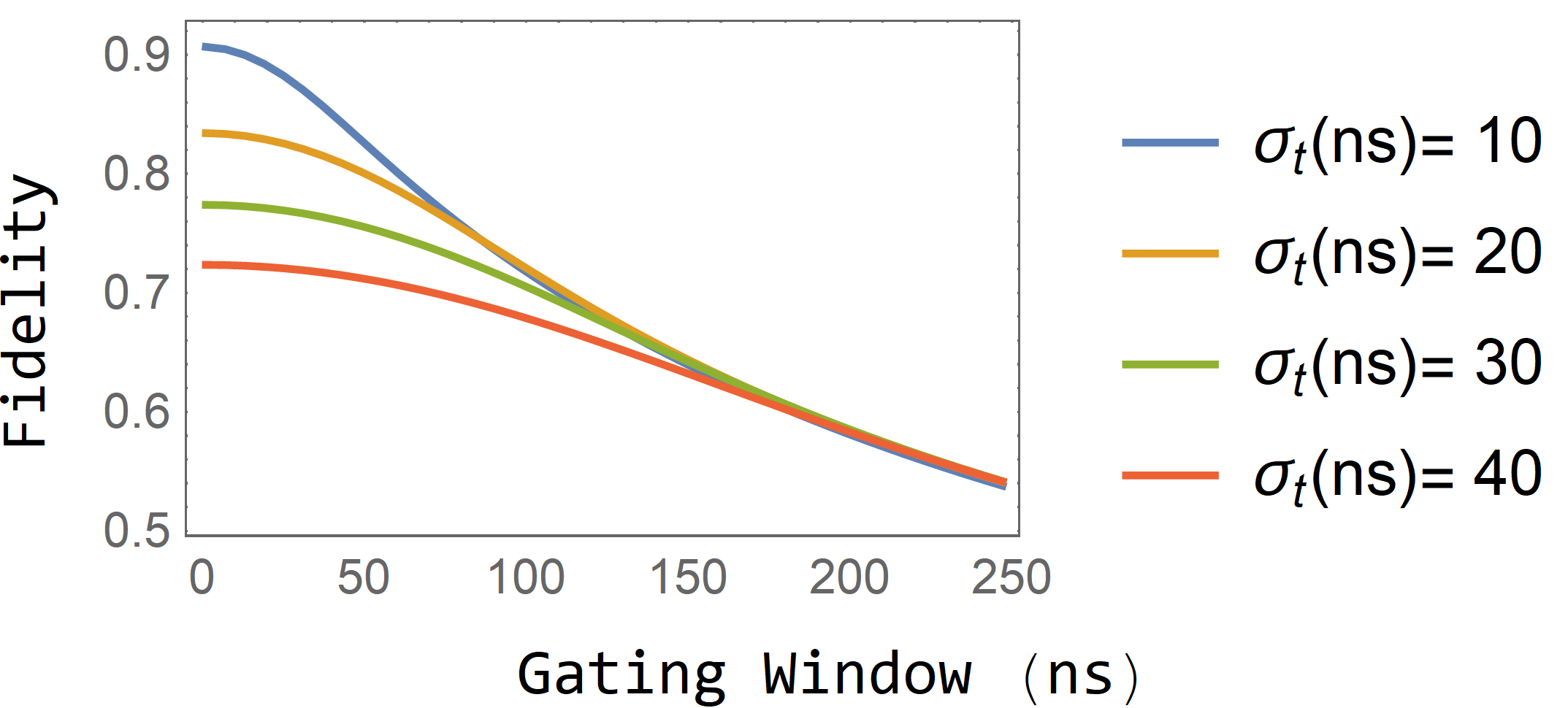}
    \includegraphics[width=0.9\linewidth]{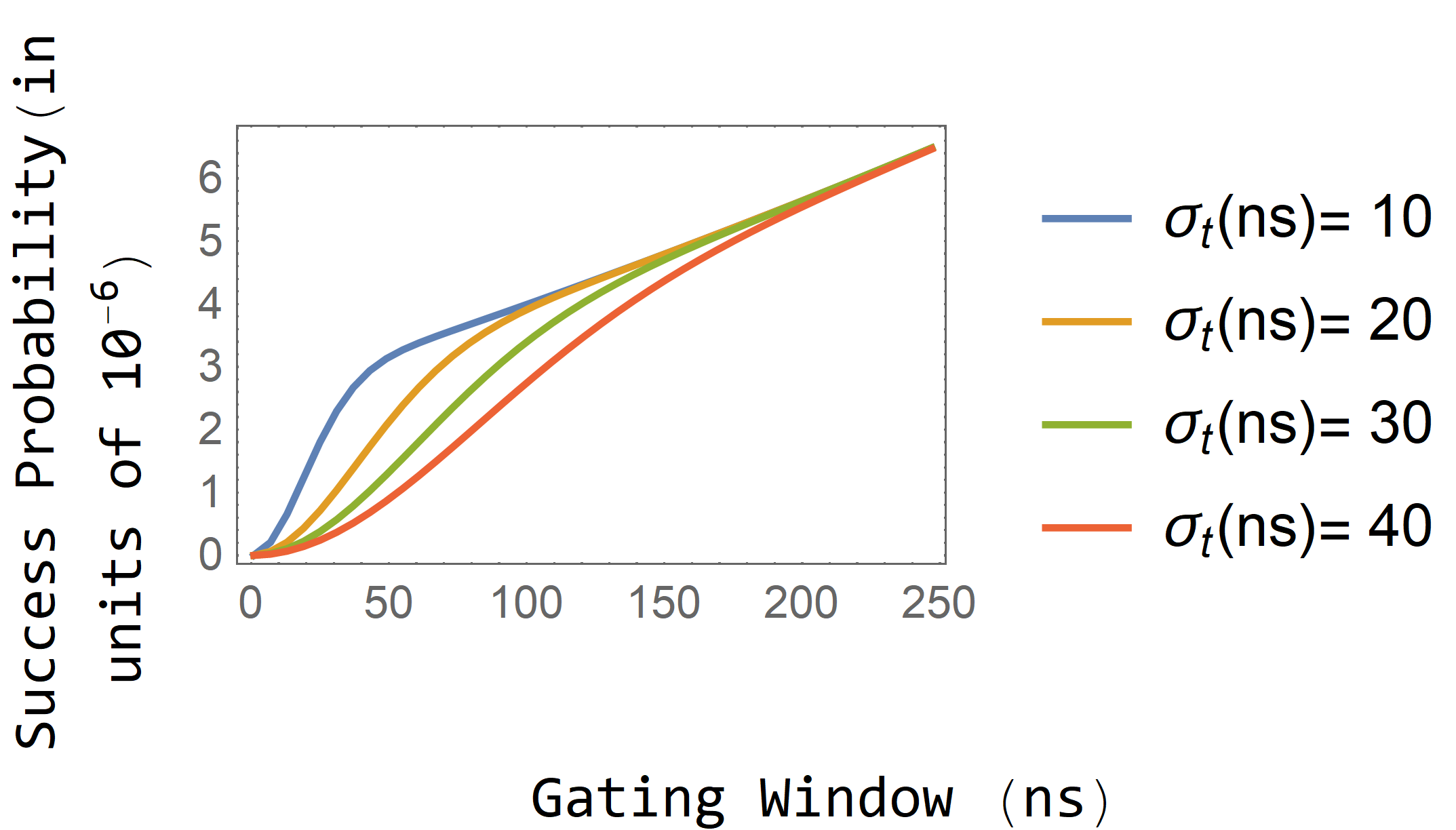}
    \caption{Fidelity $F$ and success probability $\eta_{tot}$ vs time-gating window for various wavepacket widths $\sigma_t$ (in ns). The satellite altitude is taken to be 500km and the ground stations are separated by a distance of 1000km. All other parameters are chosen according to Table 1 of the Supplementary Material}
    \label{fig:sigma_gw}
\end{figure}
\begin{figure}
    \centering
    \includegraphics[width=0.9\linewidth]{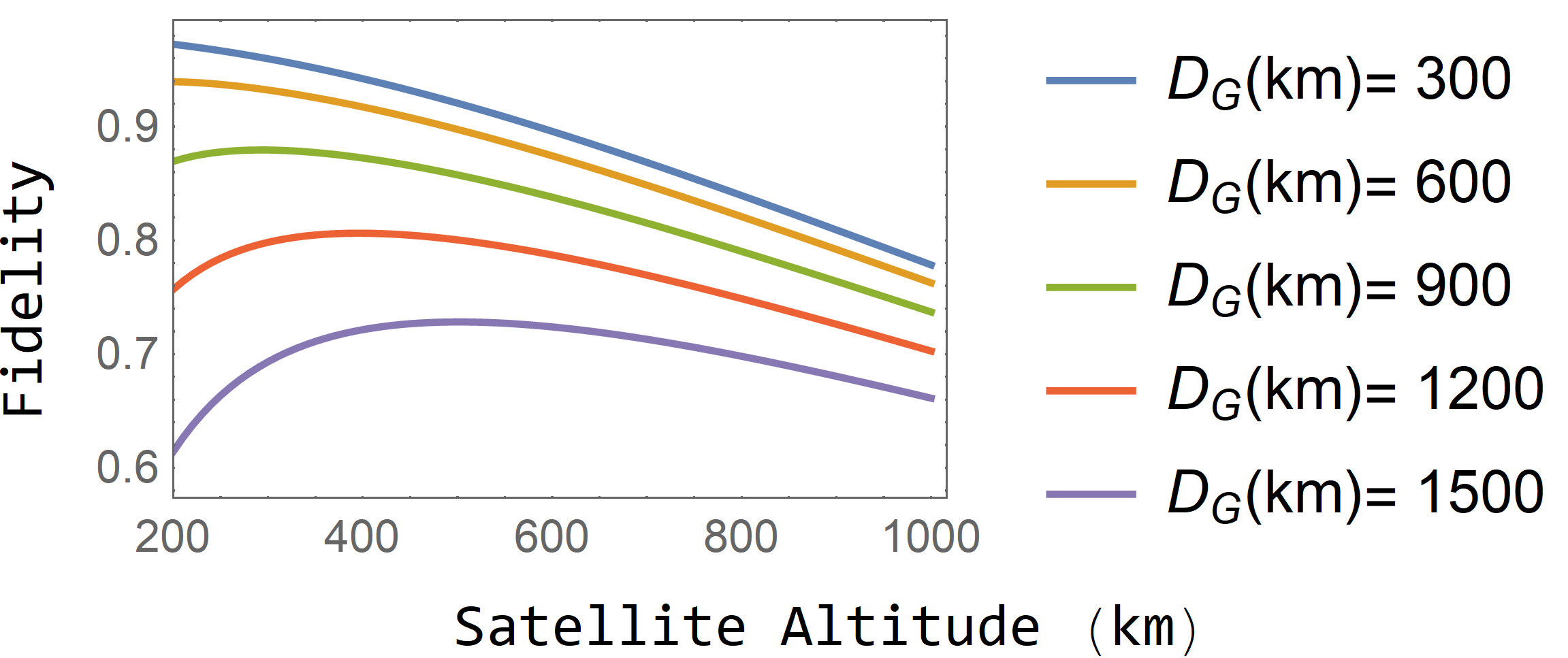}
    \includegraphics[width=0.9\linewidth]{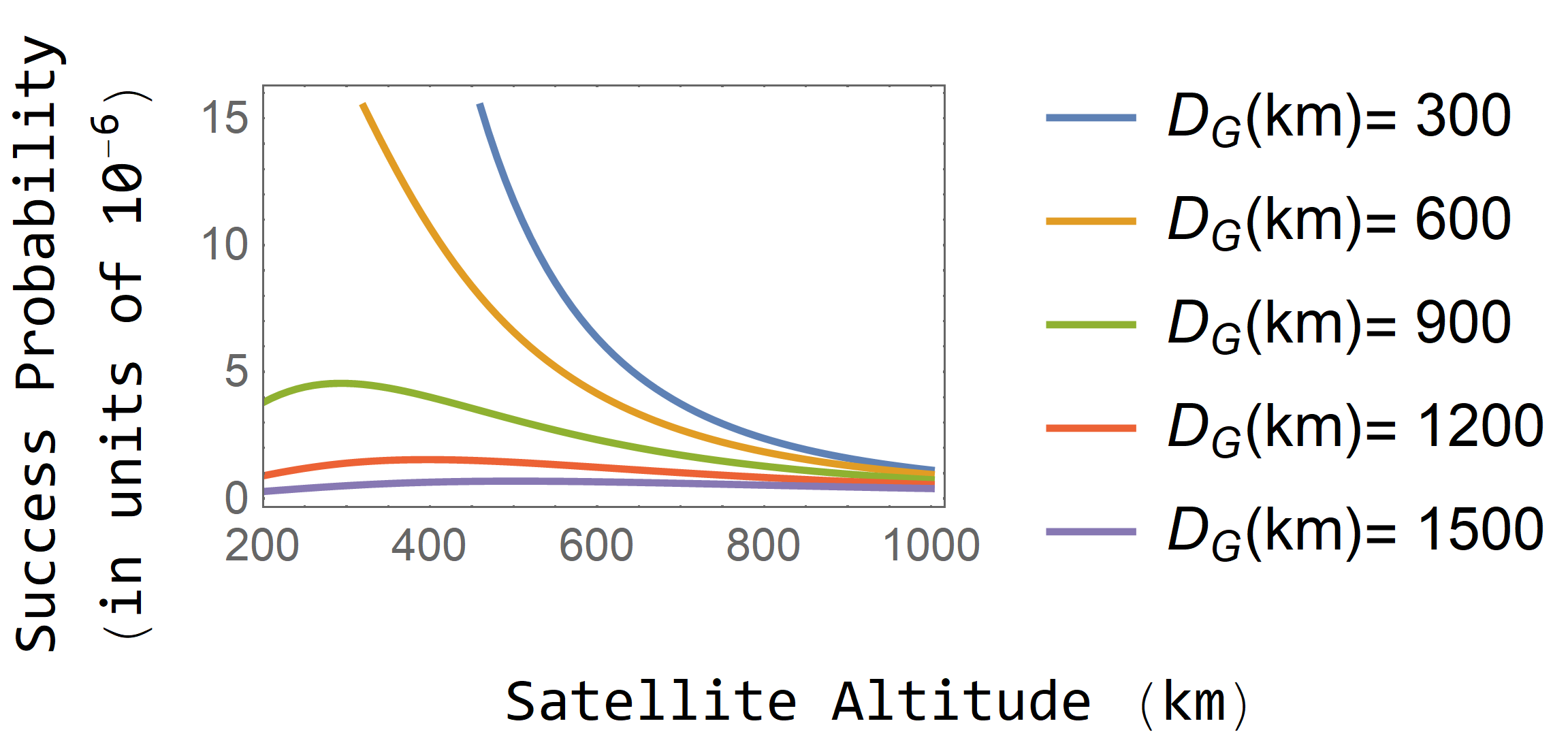}
    \includegraphics[width=0.9\linewidth]{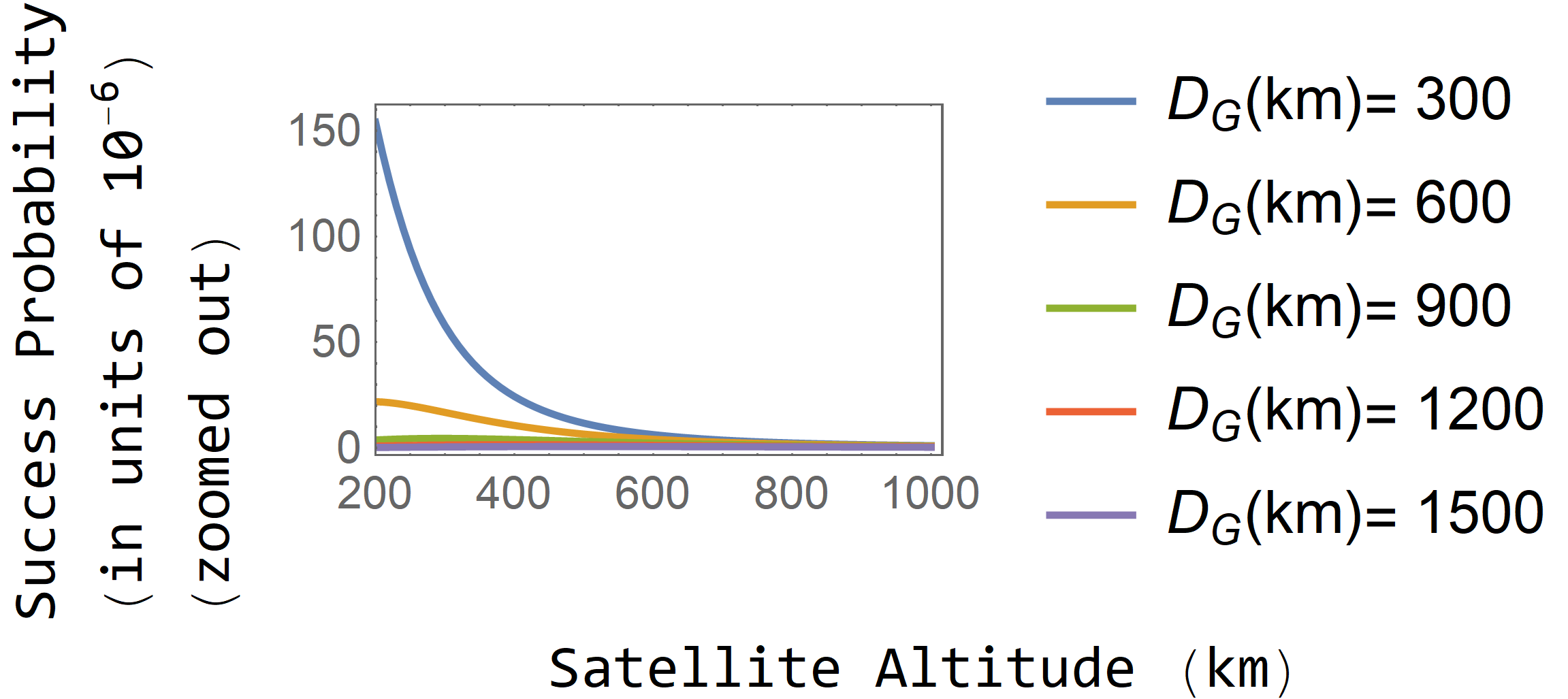}
    \caption{Fidelity $F$ and success probability $\eta_{tot}$ vs satellite altitude for various ground station separations. The third plot is the y-axis zoomed out version of the second plot, just to wholly extend the trends at $D_G=300km$ and $D_G=600km$. Here we have taken the temporal wavepacket width to be 10ns and the time gating window to be 40ns. All other parameters are chosen according to Table 1 of the Supplementary Material.}
    \label{fig:h_dg}
\end{figure}
{\em Results and Analysis.} For our analysis, we consider the snapshot of the instant where the satellite is equidistant from the two ground stations.  We are just considering the symmetry snapshot to simplify our calculations. In non-symmetric cases and cases that deviate from the symmetric snapshot, we assume that the satellite trajectory be pre-calculated and photon-release times and directions at the ground stations be appropriately adjusted to sync the photon arrival times at the satellite. This prevents any additional mode mismatch due to asymmetry. Therefore, all mode mismatch would be purely stemming from the limited precision in clock synchronization between the ground stations and the satellite.

To study how success probability~\eqref{eqn:eta_tot} and fidelity~\eqref{eqn:final_fidelity} behave, we consider realistic parameter values chosen from various experiments and industry standards [see Table 1 and Section 9 of the supplementary material]. We organize our simulation around four main quantities of interest: temporal wavepacket width $\sigma_t$, time-gating window $t = t_{\mathrm{max}} - t_{\mathrm{min}}$, satellite altitude $h$ and distance $D_G$ between ground stations. 
The first two quantities have a large range of experimentally realizable values and affect the overall fidelity and $\eta_{tot}$ non-trivially, therefore their trends must be examined to establish a reasonable range of values for these parameters. Following this, we characterize the range of distances and satellite altitudes across which the protocol can produce feasible results. 

{\em Can our uplink protocol work during daytime?} The short answer is no: our simulations show that the fidelity is around $0.25$ due to excess stray photons, which makes our protocol infeasible during daytime. From now on we focus on nighttime operation.


{\em Behavior concerning gating window and wavepacket widths.} (Figure \ref{fig:sigma_gw}) As the gating window widens, more stray photons start hitting the detector and reducing the probability of legitimate coincidences, and also the distinguishability between the two ground photon wavepackets increases, therefore decreasing the fidelity. With increasing gating window, more of the wavepackets and the stray photons are let in, which increases the overall success probability. For a given gating window, increasing the wavepacket width decreases the overall amount of wavepacket being captured by the gating window, leading to lesser fidelity and lower success probability.

{\em Behavior concerning satellite altitude and ground station separations.} (Figure \ref{fig:h_dg}) Our theoretical model predicts fidelity to decrease with satellite altitude and distance of ground stations. Initially with rising satellite altitude, the photons from the ground stations have to pass through the lesser and lesser atmosphere due to decreasing zenith angle, increasing both the fidelity and the success probability. With further increase in satellite altitude, the atmosphere begins to matter less as it becomes only a small part of the photon's journey, and the loss due to the widening of the beam's cross-sectional width $w$ starts dominating, which reduces both the fidelity and the success probability. Increasing the ground separations rapidly decreases the fidelity and success probability because with increasing ground separation, the overall photon travel paths increase, leading to longer travel, resulting in greater attenuations.

{\em Best performance.} Optimising the design of entanglement distribution in the framework of our protocol is a challenging task because the problem is a nonlinear optimization. Nevertheless, the most performant protocol we can reach is with a gating window of 40 ns, wavepacket width of 10ns, satellite altitude of 200km, and ground separation of 300km: this protocol distributes entangled state at a fidelity of 0.972 with an success probability of $1.5 \times 10^{-4}$. Nonetheless, even with a satellite altitude of 500km and a ground separation as far as 1000km, we can obtain fidelity of 0.84 and a success probability of $2.404 \times 10^{-6}$.  

\section{Conclusion}

Our analysis of uplink entanglement distribution shows feasibility only in nighttime achieving high fidelity entangled state between two ground stations with efficiencies of the order of $10^{-6}$. 
This shows promising results to motivate us to consider satellites as potential Bell measurement devices for entanglement distribution. 
We assume on-demand photon generators, but they are not experimentally of sufficient quality. Pulsed pumping generates photons probabilistically, but due to this, it is difficult to synchronize the two-photon generators at the two ground stations.
The next stage of research would be to explore this aspect and also the scalability of uplink entanglement swapping by evaluating the hardware capabilities and limitations concerning multiplexing the protocol to leverage the advantage of mammoth power capabilities on the ground as compared to the satellite. Following multiplexing, a synchronization protocol \cite{zhangtiming,zhang2024modelling} needs to be established to tag the photons for correct entanglement swapping.

\section{Acknowledgements}

The authors would like to thank Dr.Daniel Oi for their valuable comments and insights provided during their visit to UTS.

\bibliography{refs}

\end{document}